# Drift Ordered Short Mean Free Path Closure

P. J. Catto[1] and A. N. Simakov[2]
[1]*MIT Plasma Science and Fusion Center, Cambridge, MA 02139, U.S.A.*
[2]*Los Alamos National Laboratory, Los Alamos, NM 87545, USA*

**Abstract**

The complete short mean free path description of magnetized plasma in the drift ordering has recently been derived. The results correct the previous expressions for the ion pressure anisotropy (or parallel ion viscosity) and the perpendicular ion viscosity - the ion gyro-viscosity is unchanged. In addition, the electron problem is solved for the first time to obtain the electron pressure anisotropy (parallel electron viscosity) and the electron gyro-viscosity - the perpendicular electron viscosity is negligible. The results have been used to obtain a reduced fluid description appropriate to the edge of a tokamak. In addition, the complete description has been used to evaluate the radial transport of toroidal angular momentum that determines the radial electric field and flows in a tokamak.

**1. Short Mean Free Path Closure**

The short mean free path description of magnetized plasma as originally formulated by Braginskii [1] in 1957 assumes an ordering in which the ion mean flow is on the order of the ion thermal speed. Mikhailovskii and Tsypin [2] realized that this MHD ordering is not the one of most interest in many practical situations in which the flow is weaker and on the order of the ion heat flux divided by the pressure. In their ordering the ion flow velocity is allowed to be on the order of the diamagnetic drift velocity - the case of interest for most fusion devices in general, and the edge of many tokamaks in particular. Their drift ordering retains heat flow modifications to the viscosity that are missed by the MHD ordering of Braginskii. Indeed, short mean free path treatments of turbulence in magnetized plasmas must use some version of the Mikhailovskii and Tsypin results to properly retain temperature gradient terms in the gyro-viscosity. However, the truncated polynomial expansion solution technique of Mikhailovskii and Tsypin made two assumptions that we remove to obtain completely general results [3]. First, they neglected contributions to the ion viscosity that arose from the non-linear part of the collision operator. We find that removing this assumption gives rise to heat flux squared terms in the ion pressure anisotropy and perpendicular ion viscosity that are the same size as terms found by Mikhailovskii and Tsypin. Second, their truncated polynomial expansion of the ion distribution function is an inadequate approximation to the gyro-phase dependent portion of the ion distribution function. We find that their approximate form is not accurate enough to completely and correctly evaluate many of the terms in the perpendicular collisional viscosity. The modifications to the pressure anisotropy and perpendicular collisional viscosity that we evaluate are valid for turbulent and collisional transport, and also allow stronger poloidal density, temperature, and electrostatic potential variation in a tokamak than the standard Pfirsch-Schlüter ordering. We have also evaluated the electron pressure anisotropy and gyro-viscosity. Combining the ion and electron descriptions with the Maxwell equations gives a closed system of fluid equations for the plasma density, and the ion and electron temperatures and mean velocities.

## 2. Reduced Collisional Description for Tokamak Edge Plasma

Starting with our corrected short mean free path fluid equations, we derived a system of non-linear reduced moment equations, suitable for numerical modeling, that describe field-aligned fluctuations in low-beta collisional magnetized edge plasma [4]. These equations advance the plasma density, electron and ion pressures (or, equivalently, temperatures), parallel ion flow velocity, parallel current, vorticity (or, equivalently, electrostatic potential), perturbed parallel electromagnetic potential, and perturbed magnetic field. The equations locally conserve particle number and total energy, and insure that perturbed magnetic field and total plasma current are divergence-free. In addition, while intended primarily for modeling plasma edge turbulence, they contain the neoclassical results for plasma current, parallel ion flow velocity, and parallel gradients of equilibrium electron and ion temperatures. These equations assume that neoclassical transport of angular momentum is unimportant since they employ the gyro-viscous cancellation (which assumes that the variation of the magnetic field is weak compared to the spatial variations of density and temperature). Consequently, they assume that the turbulence dominates over neoclassical transport to set the radial electric field. However, more work is necessary to determine whether the assumptions that go into deriving reduced descriptions in general and the vorticity equation in particular are valid because the approximations employed can introduce spurious electric fields. These spurious radial electric fields can be removed by subtracting off the appropriate flux surface averages that arise because of the use of the gyro-viscous cancellation and other approximations. It is important to remember that in the absence of turbulence and ion temperature variation a radial Maxwell-Boltzmann, rigid rotor response $e\partial\Phi/\partial\psi + n^{-1}\partial p_i/\partial\psi = $ constant must be obtained, where n, $p_i$ and $\Phi$ are the plasma density, ion pressure, and electrostatic potential, respectively, with $\psi$ the poloidal flux function and e the magnitude of the charge on an electron (we assume singly charged ions).

## 3. Angular Momentum Transport in the Pfirsch-Schlüter Regime

For neoclassical transport in general tokamak geometry the radial electric field is determined by the condition that the radial flux of toroidal angular momentum vanish, that is, by $\langle R^2\nabla\zeta\cdot\vec{\vec{\pi}}\cdot\nabla\psi\rangle = 0$ where $\vec{\vec{\pi}}$ is the ion stress tensor, $\zeta$ is the toroidal angle, R is the cylindrical radial distance from the symmetry axis, and $\psi$ is the flux function associated with the magnetic field $\vec{B} = I\nabla\zeta + \nabla\zeta\times\nabla\psi$. In a collisional tokamak plasma this neoclassical limit is referred to as the Pfirsch-Schlüter regime since they were the first to investigate the return particle and heat flows that are necessary to satisfy the lowest order particle and energy balance equations. The general expression for the radial flux of toroidal angular momentum is quite complex and the first systematic evaluation was by Hazeltine in 1974 [5]. Our results [6] differ from his for the following two reasons: (i) his expression for the radial flux of toroidal angular momentum is incomplete [7] - he solved a kinetic equation [8] that can be shown to be missing some second order in the ion gyro-radius expansion terms needed to obtain the full gyro-viscosity [2, 3] as well all higher order terms needed for a direct determination of the perpendicular viscosity, and (ii) he assumed that both ion pressure and electrostatic potential separately had no poloidal variation rather than requiring that they need only satisfy the constraint of parallel ion momentum conservation.

Although the general expression for the radial electric field is quite lengthy [6], two simple limits can be deduced: (i) the limit of concentric circular flux surfaces and (ii) the case of a strong up-down asymmetry as might be expected just inside the separatrix in single null divertor geometry. The general expressions are substantially more involved because the gyro-viscosity must be evaluated to higher order in the Pfirsch-Schlüter expansion procedure that assumes $\delta \equiv \rho/L_\perp \ll \Delta \equiv \lambda/L_\parallel \ll 1$, with $\rho$ and $\lambda$ the ion gyro-radius and ion mean free path and $L_\perp$ and $L_\parallel$ the perpendicular and parallel scale lengths.

We assume that the plasma current $I_p$ is in the direction of increasing toroidal angle (the $\nabla\zeta$ direction) in order to make $\psi$ increase outward from the magnetic axis. As a result, the direction of the toroidal magnetic field is determined by the sign of the flux function I. The curvature and gradient B drift are towards a lower X-point when I is positive and $I_p$ in the $\nabla\zeta$ direction. To write down our results in a compact form it is convenient to define a rotation frequency $\omega = -c[\partial\Phi/\partial\psi + (en)^{-1}\partial p_i/\partial\psi]$ since in the absence of temperature variation and momentum sources or sinks the only solution allowed is one that is $\partial\omega/\partial\psi = 0$, which is consistent with the generalized radial Maxwell-Boltzmann response $\omega$ = constant.

For an up-down symmetric tokamak, the lowest order gyro-viscous contribution to the radial flux of toroidal angular momentum vanishes, and the next order correction in the Pfirsch-Schlüter expansion ($\delta/\Delta \ll 1$) must be evaluated. The resulting expression for arbitrary cross section, aspect ratio, magnetic field, and plasma pressure simplifies substantially for a circular, concentric flux surface model. For this case it is convenient to denote the radius of the flux surface by r and use $B_t = B_0 R_0/R$ for the toroidal magnetic field with $R_0$ the radius of the magnetic axis and $r/R_0 \ll 1$. We then find that the shear in the frequency $\omega$ is simply given by

$$\frac{r}{\Omega_0}\frac{d\omega}{dr} \approx -\frac{0.19 q^3 \rho_0^2 T_e}{T_e + T_i}\left(\frac{d\ln T_i}{dr}\right)^2, \qquad (1)$$

where q is the safety factor and we define $\Omega_0 = eB_0/Mc$ and $\rho_0 = v_i/\Omega_0$, with $v_i = (2T_i/M)^{1/2}$ the ion thermal speed and M the ion mass. The result of Eq. (1) is in agreement with the result of Claassen and Gerhauser for $T_e = T_i$ [9]. According to (1) the radial variation of ion temperature is responsible for driving a shear in the electric field that results in a departure from radial Maxwell-Boltzmann behavior. Consequently, the ion flow is sheared as well, and to lowest order can be written in the form

$$\vec{V} = \omega(\psi)R^2\nabla\zeta + u(\psi)\vec{B}, \qquad (2)$$

where $u \approx -(1.8cI/e\langle B^2\rangle)\partial T_i/\partial\psi$ [5]. Notice that the shear in the poloidal flow is controlled by the ion temperature gradient rather than the radial electric field

For the strongly up-down asymmetric case of a single null divertor the expression for the shear in the electric field (or $\omega$) in general tokamak geometry is given by

$$\frac{d\omega}{d\psi} \approx -\frac{4IdT_i/d\psi}{3M\nu\langle B^2\rangle}\frac{\langle R^2\vec{B}\cdot\nabla\ell nB\rangle}{\langle R^2B_p^2B^{-4}(R^2B^2+3I^2)\rangle}, \qquad (3)$$

where $\langle R^2\vec{B}\cdot\nabla\ell nB\rangle$ vanishes for an up-down symmetric configurations and determines the sign of the shear $d\omega/d\psi$ for asymmetric ones. Again, once (3) is solved for $\omega$ it can be inserted into (2) to find the ion flow. Expression (3) depends on the ion collision frequency $\nu$ since it is found by balancing the lowest order gyro-viscosity with the collisional perpendicular viscosity. The up-down symmetric result of (1) is obtained by evaluating the gyro-viscosity to higher order in the $\delta/\Delta << 1$ expansion so the collision frequency cancels out and then performing an aspect ratio expansion. It takes extremely strong up-down asymmetry to make the asymmetric term on the right side of (3) dominate over the symmetric term. Normally the general expression with both types of terms retained must be employed for single null divertor configurations.

## 4. Drift Kinetics

The drift kinetic equation of Hazeltine [8] is widely viewed as the best available. However, we have recently shown that it does not contain information needed to use it to evaluate the full gyro-viscosity [10]. The reason is that the Hazeltine derivation assumes that the magnetic moment dependence and the energy dependence of the distribution function are both the same order. In most magnetically confined plasmas the lowest order distribution function is isotropic in velocity space. If the Hazeltine drift kinetic equation is used in such situations then it is missing some terms that are second order in the gyro-radius expansion. When these terms are retained the equation becomes more complex, but the full gyro-viscosity can be obtained for arbitrary collisionality [10].

## 5. Summary

We have performed an in depth study of collisional plasmas and applied the results to tokamaks to obtain a reduced description and to evaluate the axisymmetric neoclassical Pfirsch-Schlüter radial electric field for arbitrary cross-section and pressure. If, as believed, energy inverse cascades from short turbulent scales to large structures which set-up axisymmetric sheared zonal flows that control the turbulence level, then these neoclassical features must be retained in a complete description to evaluate the full axisymmetric response.

**Acknowledgments**
Research supported by U.S. DoE by grants DE-FG02-91ER-54109 at MIT and W-7405-ENG-36 at Los Alamos National Laboratory.